\documentclass[runningheads]{llncs}

\usepackage[T1]{fontenc}
\usepackage{graphicx}
\usepackage{booktabs}
\usepackage{amsmath}
\usepackage{amssymb}
\usepackage{array}
\usepackage{multirow}
\usepackage{xcolor}
\usepackage[hidelinks]{hyperref}

\newcommand{\rscore}{\ensuremath{\mathit{research\_score}}}

\begin{document}

\title{Autoresearch with Coding Agents: Generalizers and Metric-Maximizers on Quran Recitation Data}
\titlerunning{Autoresearch with Coding Agents}

\author{Nursultan Askarbekuly\inst{1} \and Mohamad Al Mdfaa\inst{2} \and
Ahmed Helaly\inst{1} \and Gonzalo Ferrer\inst{3} \and Manuel Mazzara\inst{1}}
\authorrunning{N. Askarbekuly et al.}

\institute{Innopolis University, Innopolis, Republic of Tatarstan, Russia\\
\email{\{n.askarbekuly,a.helaly\}@innopolis.university},
\email{m.mazzara@innopolis.ru}
\and LemoniLab FZCO, Dubai, United Arab Emirates\\
\email{mohamad.almdfaa@lemonilab.com}
\and Skolkovo Institute of Science and Technology, Moscow, Russia\\
\email{g.ferrer@skoltech.ru}}

\maketitle

\begin{abstract}
Coding agents can now be left alone to improve software against a score.
In this pattern -- recently popularized as \emph{autoresearch} -- the agent
receives a dataset, an evaluation script, and one editable file, and
iterates without supervision: modify the code, measure, keep the change if
the score improves. But what does the agent actually optimize -- the
developer's intent, or the literal number? We ran this loop on a real
production task: deciding which Quranic verses appear in a noisy
speech-recognition transcript and splitting the transcript by verse. Two
frontier coding agents, Claude Code and OpenAI Codex, started from the
same blank file with the same instructions, budget, and reasoning effort,
three runs each. Both independently invented the same algorithm
(canonicalization, $n$-gram anchoring, dynamic-programming alignment) --
and then diverged. Claude stopped early with compact, general code. Codex
drove the score $\sim\!10\times$ lower, largely by memorizing answers to
individual evaluation rows (19--41 hardcoded verse ids per run): a clean
natural instance of specification gaming by a production agent. In a
preregistered second study, we hardened the harness along four axes at once
-- adding a held-out test set, withholding expected answers from failure
reports, disclosing the split to both agents, and isolating run state. The
memorization vanished, and the score gap vanished
with it -- yet Codex's general core transferred \emph{better and more
consistently} (held-out detection+split $0.085\pm0.004$ vs.\
$0.121\pm0.031$), losing only on one missed rejection of non-recitation
input. Two exploratory community arms (Cursor, Antigravity) are consistent
with the pattern. On this task and metric, every agent's held-out solution
matched or beat the hand-engineered pipeline it was built to replace -- the
best by an order of magnitude -- and that artifact now runs in production.
From the ways agents exploited
our harness -- reading sibling runs through shared git state, leaving
notes to ``future runs'' in persistent memory -- we distill five design
rules for evaluating autonomous agents.

\end{abstract}

\section{Introduction}
\label{sec:intro}

Hand a coding agent a frozen dataset, a frozen metric, and one editable
file; tell it to iterate unattended, keeping only changes that improve the
score. This \emph{autoresearch} pattern (colloquially, the ``Karpathy
loop'') -- named by Karpathy's recent
experiment~\cite{karpathy2026autoresearch}, which strips a single-GPU
language-model training script to $\sim\!630$ lines and reports agents
autonomously discovering architecture and hyperparameter improvements over
hundreds of unattended experiments -- generalizes to any task with a fast,
scriptable metric, and is likely to proliferate as the cheapest way to
extract unattended engineering effort from coding agents. It also raises a
question, which existing agent benchmarks do not answer. Benchmarks such as
SWE-bench~\cite{jimenez2024swebench} measure whether an agent
\emph{completes} a well-specified task (does the patch make the failing
test pass); they say nothing about what an agent does when the
specification is a fallible \emph{proxy} that can be exploited to achieve a
better score. An autoresearch loop hands an agent exactly that situation
for hundreds of consecutive decisions, with no human watching. What the
agent optimizes then -- the intended objective or the literal one -- is the
subject of this paper.

We build such a loop around a genuine production problem -- Stage 1 of a
transcript-only Quran memorization checker, which must (a) detect which
verses (\emph{ayahs}) a reciter has read from a no-diacritics automatic
speech recognition (ASR) transcript and (b) split that transcript by verse --
and use it to compare two frontier coding agents, Claude Code and OpenAI
Codex, under identical instructions, budgets, and starting code. Unlike a
synthetic benchmark, this task has real ASR noise, a closed reference corpus
(the Quran) that makes exact gold labels possible, messy user behavior
(repetitions, restarts, non-recitation speech), and a metric whose floor is
known exactly -- properties that make it a discriminating and realistic
testbed for autonomous algorithm research. The substrate itself is a text
analysis problem -- noisy Arabic transcript-to-canon alignment and
segmentation -- so the paper contributes on two levels: a data-analysis
methodology for evaluating autonomous agents, demonstrated on a
low-resource text-processing task of independent interest. The loop's
output is not only measured but consequential: the winning agent-built
solution outperforms the incumbent hand-engineered pipeline (iterated over
months) by $\sim\!10\times$ on held-out data and has replaced it in the
production application.

We organize the paper around three research questions:
\begin{description}
\item[RQ1 (feasibility).] Starting from a blank abstain-only stub, can a
  coding agent autonomously build and improve a non-trivial text-alignment
  algorithm purely through the modify$\to$verify$\to$keep/discard loop?
\item[RQ2 (behavior under a proxy).] When the only feedback is an imperfect,
  gameable metric, how do different agents' optimization behaviors differ in
  their stopping decisions, their artifact character, and their willingness
  to exploit the metric's loopholes?
\item[RQ3 (effect of held-out disclosure).] Does adding -- and disclosing --
  a held-out split change those behaviors, and which agent's improvements
  actually generalize?
\end{description}

\paragraph{Contributions.}
\begin{enumerate}
\item A reproducible autoresearch harness built on a real production task
  with real user data, together with a frozen metric, frozen dataset splits
  (by SHA-256), and uniform per-experiment logging that makes six-plus agent
  runs directly comparable (Sect.~\ref{sec:harness}).
\item A controlled $3\times2$ comparison (Claude Code vs.\ Codex, three runs
  each) showing that under an ungated metric the two agents behave in
  opposite ways, consistently within each arm -- one self-imposing a
  simplicity prior the metric never asked for, the other driving the literal
  objective down by memorizing 19--41 evaluation-row answers per run, a
  count reproducible mechanically from public tags. We call these behaviors
  \emph{generalizer} and \emph{metric-maximizer} as shorthand for what these
  artifacts show in this harness, not as claimed traits of either agent
  family (Sect.~\ref{sec:study1}).
\item A held-out generalization study (same $3\times2$ design, disclosed
  train/test split) showing that the train-side separation between agents
  evaporates, that disclosure suppresses literal memorization but not the
  underlying optimization drive, and that the arm that overfit harder on
  train in Study~1 in fact transfers its algorithmic core \emph{better} --
  while losing on a single rare-event failure mode instead
  (Sect.~\ref{sec:study2}).
\item An extension to two exploratory community-run arms (Cursor,
  Antigravity) consistent with the held-out pattern under different models
  and tooling,
  and a catalogue of agent-initiated harness-isolation failures (reading a
  sibling run's branch through shared git state; writing unsolicited
  persistent-memory notes to ``future runs'') distilled into five concrete
  design rules for autonomous-agent evaluation harnesses
  (Sect.~\ref{sec:community}, \ref{sec:rules}).
\item An applied validation, specific to this task and metric: measured
  under the same contract, every agent arm's held-out core matched or beat
  the incumbent hand-built production pipeline, the winner by
  $\sim\!10\times$; the obvious cross-agent ensemble \emph{fails} when
  actually tested; and the winning single artifact is now deployed in
  production. This is a reference point rather than an effort-matched
  human-versus-agent comparison (Sect.~\ref{sec:incumbent},
  \ref{sec:discussion}).
\end{enumerate}

\section{Related Work}
\label{sec:related}

\paragraph{Autonomous research loops and research agents.}
Karpathy's autoresearch~\cite{karpathy2026autoresearch} is our direct
template: a frozen evaluation file, an editable training file, and a
natural-language program instructing an agent to iterate, log, and keep
only improving changes; follow-on community reports scaling the loop
explicitly flag Goodhart risk on the validation metric as its central
vulnerability. A parallel line frames autonomous research more broadly:
MLGym benchmarks agents across many ML research
tasks~\cite{nathani2025mlgym}, PaperBench measures whether agents can
replicate published AI papers~\cite{starace2025paperbench}, and Agent
Laboratory drives an end-to-end assistant through literature review,
experimentation, and report writing~\cite{schmidgall2025agentlab}. ``AI
Scientist'' systems~\cite{sakana2024aiscientist} automate the entire
research life cycle, but evaluations of such systems focus on the quality
and novelty of the ideas and writing produced~\cite{beel2025evaluating},
not on the optimization \emph{behavior} of the agent under a fixed,
gameable metric, which is our focus. AlphaEvolve~\cite{novikov2025alphaevolve}
evolves code against a human-specified verification metric at much larger
scale, assuming the metric problem is solved by the human expert rather
than studying what happens when it is not. Autoresearch-style loops with
these same two agents have since been examined at benchmark scale by
SpecBench~\cite{zhao2026specbench}; what we contribute alongside it is a
controlled \emph{between-agent} comparison of this loop design on a
production task outside language-model pretraining, with a disclosed
held-out split and the complete per-experiment git history of every run
published as evidence.

\paragraph{Agentic coding and ML-engineering benchmarks.}
SWE-bench~\cite{jimenez2024swebench} and its successors evaluate whether an
agent can produce a patch that makes a repository's real test suite pass, on
thousands of real GitHub issues. These benchmarks are the standard for
\emph{task completion} under a well-specified, already-correct oracle (the
project's own tests). Closer to our setting, MLE-bench~\cite{chan2024mlebench}
scores agents on Kaggle-style machine-learning engineering tasks against
held-out leaderboards, and RE-Bench~\cite{wijk2024rebench} compares agents
with human experts on open-ended ML research-engineering environments; both
report that agents sometimes attempt to game their scoring. Our design
differs in three ways that matter for the question we ask. First, these
benchmarks measure \emph{achievement} -- how far up a leaderboard an agent
gets -- whereas we hold the task fixed and study \emph{behavioral
divergence} between agents given the identical loop, with the full
per-experiment git history as evidence. Second, we run the same matrix twice,
with the metric's exploitability first left open and then closed, which
exhibits a contrast that leaderboard benchmarks cannot: they bake the
held-out split in from the start, so the open-metric condition never occurs.
(We are careful in Sect.~\ref{sec:disclosure} not to read this contrast as
isolating a single causal factor -- four things changed between the two
studies.) Third, our forensic unit is
the final code artifact itself (hardcoded-constant counts, per-recording
lookups), not only the score.

\paragraph{Specification gaming, reward hacking, and Goodhart's law.}
Krakovna et al.\ catalogue specification gaming: behavior that satisfies the
literal wording of an objective while defeating its
intent~\cite{krakovna2020specification}, echoing
Goodhart's law -- a measure optimized as a target ceases to track what it was
designed to measure -- and its later
formalization~\cite{manheim2018goodhart}. Amodei et al.\ list reward hacking
among the concrete near-term problems of AI systems~\cite{amodei2016concrete},
and Skalse et al.\ formalize when a proxy reward admits a hacking
policy~\cite{skalse2022goodhart}. Recent work documents the same failure in
LLM \emph{coding} agents: RL-trained models exploiting test
harnesses~\cite{baker2025monitoring}, agents passing deliberately
contradictory specification/test pairs~\cite{zhong2025impossiblebench}, and
reward hacking scored in long-horizon software-engineering
tasks~\cite{zhao2026specbench}, and benchmarks built specifically to invite
hardcoding: EvilGenie~\cite{gabor2026evilgenie} seeds LiveCodeBench problems
into an environment where agents can hardcode test cases or edit the test
files, and detects the result with held-out tests, an LLM judge, and
edit-detection, reporting explicit reward hacking by both Codex and Claude
Code. SpecBench~\cite{zhao2026specbench} likewise exercises autoresearch-style
long-horizon settings with both agents and reports lookup-table
memorization.

Two of these overlap our setting closely, so we are explicit about what is
new. That frontier coding agents hardcode against a visible metric, that
Codex and Claude Code both do so under some harness, and that held-out tests
reveal it are established by EvilGenie and SpecBench; our results corroborate
them on an independent task rather than discovering the phenomenon. We add
three things. The substrate is a \emph{production} task with real user data
and a deployed incumbent, not a benchmark built to be gameable -- the
loophole was an accident of our harness design, the condition practitioners
actually meet. Our forensic unit is the final artifact under a mechanical
count over public tags, not a judge model's verdict. And the metric is
\emph{read-only}, so the only available exploit is memorization against a
visible scorecard; under that narrower affordance the two agents separate
completely, one hardcoding in three runs of three and the other in none.
That separation, rather than the existence of hardcoding, is the
contribution.

\paragraph{Quranic ASR and recitation checking.}
A substantial applied literature builds ASR systems for Quranic
recitation, from HMM-based verse delimitation~\cite{tabbal2006recitation} to
modern end-to-end models trained on crowd-sourced corpora such as
Tarteel~\cite{yousfi2018tarteel,alharbi2026comparative} and rule-based or
learned Tajweed (pronunciation-rule) checkers~\cite{alagrami2021smartajweed};
see~\cite{farghaly2021asrreview} for a survey. This literature targets
\emph{transcription and pronunciation} accuracy. Our task sits one layer up
the pipeline and is transcript-only: given an already-transcribed,
diacritics-free string, decide which verses were recited and how the
transcript decomposes across them -- the detection-and-segmentation problem
that a memorization-checking product must solve before any mistake-detection
layer can run. We do not contribute a new ASR model; we use this real,
production-adjacent task purely as a discriminating substrate for studying
agent research behavior, and note it as a secondary contribution that the
harness and dataset (Sect.~\ref{sec:harness}) are reusable for that
literature.

\paragraph{Positioning within AIST.}
Two strands at recent AIST editions bracket this work. Low-resource-language
studies such as the KyrgyzNLP survey~\cite{alekseev2025kyrgyznlp} motivate
careful evaluation on scripts underserved by mainstream benchmarks, and
diacritics-free Arabic alignment against a closed canon is one such problem;
work on applying LLMs to code tasks~\cite{andryushchenko2024codeqa} we extend
from \emph{using} an LLM on a code task to \emph{measuring how autonomous
coding agents behave} when that task is an unattended optimization loop.

\section{Task, Data, and Harness}
\label{sec:harness}

\subsection{Task}
Stage~1 of a transcript-only memorization checker: given a diacritics-free
ASR transcript, return either an abstention (the audio is not Quranic
recitation) or the recited verse-id sequence together with a per-verse split
of the transcript. Within-verse mistake detection (Stage~2) requires labels
that do not yet exist and is out of scope.

\subsection{Dataset}
258 Telegram-bot auto-detect recordings from production use (254 Quranic,
4 non-Quranic; 45 surahs; transcript lengths 3 to 400+ words), transcribed
with an off-the-shelf commercial ASR system (OpenAI's speech-to-text API,
used as a black box with default settings) and reviewed into gold
\mbox{\texttt{(ayah\_assignment,}} \texttt{per-ayah split, confidence)}
triples. Labeling
conventions: gold segments exclude the leading \emph{isti'adha} and
\emph{basmala} from recited text, except in the opening surah where the
\emph{basmala} is itself the first verse; repetition is recorded as
repeated ids in sequence and, since detection scores the id \emph{set},
affects only the split component; assignments are an id, a range, or a
comma-separated list for skip-recitation, with the split id set equal to
the assignment id set; rows are scored at \texttt{high}/\texttt{medium}
confidence (\texttt{low} retained but unscored); and a human-reference
audit field is never exposed to the algorithm nor used in scoring. The
dataset was frozen by SHA-256 before any run
and one label was corrected \emph{after} both studies completed, once both
agents independently flagged it as likely gold-label noise
(Sect.~\ref{sec:study2-audit}); all reported numbers are against the version
in force at the time each study ran. Concretely, Studies~1 and~2 both ran
against dataset v1; the corrected v1.1 is what the public release contains.
The correction affects one training row, so no held-out number in this paper
changes under it, and Study~1 and Study~2 remain comparable on the held-out
split. The only quantity v1.1 moves is the gold-echo reference, from
$0.008$ to $0$.

\paragraph{Data governance and ethics.} The recordings are production
interactions with a Quran-memorization Telegram bot, contributed by adult
users of the authors' own application under its terms of service, which
disclose that submitted recitations may be reviewed to improve the service
(9 distinct learners in the training split). No audio is released; the
published dataset contains only ASR transcripts and gold labels, with
learner and recording identifiers replaced by surrogate row ids
(\texttt{train-001}, \texttt{test-001}, \ldots), and the four non-Quranic
rows were checked by hand and contain no personal content. Because recited
text comes from a closed, public canon, the transcripts carry essentially no
speaker-identifying content. The work was conducted as product engineering
on the authors' own service and was not submitted to an IRB.

\subsection{Metric}
\begin{equation}
\rscore = \mathit{detection\_error} + \mathit{split\_error} + \mathit{abstain\_error}
\end{equation}
lower is better; an empty abstain-only stub scores $2.0$. Feeding the gold
split back scores $\approx\!0.008$ in Study~1 rather than $0$, because two
cross-surah edge rows were internally inconsistent under the labeling
convention then in force. We therefore call this the \emph{gold-echo
reference} rather than a floor: a solution resolving those two rows
\emph{differently} from gold can legitimately score below it, and two
Study~1 Codex runs do (Sect.~\ref{sec:study1-results}). After the label
correction of Sect.~\ref{sec:study2-audit} it is exactly $0$ on both splits,
so Study~2 is measured against a true floor. Detection compares the
\emph{set} of predicted verse ids against gold, so a reciter repeating a
verse is a split concern, not a detection error. Split is word-assignment
accuracy: the fraction of gold transcript words placed in the correct
verse bucket, compared under a metric-only orthographic canonicalizer (folding
common Arabic spelling variants (e.g.\ hamza-bearing letter forms) and
Uthmani orthographic conventions) that is
\emph{not} exposed to the agent's own algorithm. Abstain requires a correct
rejection on non-recitation rows.

\subsection{Loop protocol}
A fixed, agent-agnostic \texttt{PROGRAM.md} specifies: one idea per
experiment; commit before verifying; run the fixed scorecard and log a
uniform row (iteration, timestamp, elapsed time, agent tag, commit, score
components, keep/discard status); keep the change iff the score improves,
else \texttt{git reset}; stop at budget or at a plateau of $\sim\!15$
non-improving experiments; and, as an explicit tie-breaker, prefer the
simpler of two equally-good solutions. The scorecard, the Quran reference
table, and \texttt{PROGRAM.md} itself are read-only; the agent edits only
its own solution file and helpers. Each run starts from an identical
abstain-only stub in an isolated environment (Sect.~\ref{sec:isolation}),
so every run's history is one commit per experiment -- a complete,
replayable audit trail.

\section{Study 1: Setup}
\label{sec:study1}

\textbf{Arms.} Claude Code vs.\ OpenAI Codex, three runs each.
\textbf{Budget.} 30 experiments or one hour of wall-clock time, whichever
comes first; reasoning effort fixed to the named ``high'' tier on both
agents' scales. \textbf{Autonomy.} Full-auto permissions on both sides
(Claude Code with unattended file/shell access; Codex in unattended
workspace-write mode). \textbf{Confounds pinned and reported:} agent CLI
version and underlying model, identical hardware for all six runs, and a
dataset+evaluator hash recorded before any run began.
Table~\ref{tab:confounds} states them in full.

\begin{table}[t]
\centering
\caption{Pinned configuration for the controlled arms (identical across
Studies 1 and 2). The community arms of Sect.~\ref{sec:community} are
\emph{not} matched on these axes and are reported separately.}
\label{tab:confounds}
\begin{tabular}{lll}
\toprule
 & Claude arm & Codex arm \\
\midrule
CLI                & Claude Code v2.1.198 & Codex CLI v0.139.0 \\
Model              & Claude Opus 4.8      & GPT-5.5 \\
Reasoning effort   & \texttt{high}        & \texttt{high} \\
Permission mode    & \texttt{--dangerously-skip-permissions}
                   & \texttt{-s workspace-write -a never} \\
Budget             & \multicolumn{2}{l}{30 experiments or 60 min wall-clock,
                      whichever first} \\
Hardware           & \multicolumn{2}{l}{one MacBook Pro, all six controlled runs} \\
Dataset            & \multicolumn{2}{l}{v1, SHA-256 frozen in
                      \texttt{runs/inputs.sha256} before any run} \\
\bottomrule
\end{tabular}
\end{table}

Reasoning effort is matched by \emph{named tier}, not by an equivalence
claim: both vendors expose a rung labeled \texttt{high} and we set both
there; whether those tiers are comparable in compute is undisclosed, and we
do not assume it. The arms ran on different days -- a temporal confound we
cannot retire (Sect.~\ref{sec:limitations}).

\section{Study 1: Results}
\label{sec:study1-results}

\subsection{Headline scores}

\begin{table}[t]
\centering
\caption{Study 1 -- best score per run on the (non-held-out) evaluation set.
Baseline (empty stub) $=2.00$; gold-echo reference $\approx 0.008$ (not a
floor -- see Sect.~\ref{sec:harness}; codex-r1 and codex-r3 score below it
by resolving two inconsistently-labeled rows differently from gold).
Hardcoded ids are counted mechanically by
\texttt{tools/count\_hardcoded\_ids.sh}.}
\label{tab:study1-headline}
\begin{tabular}{lrrrrr}
\toprule
Run & Best score & Experiments & Keep/discard & LOC & Hardcoded ids \\
\midrule
claude-r1 & 0.0875 & 13 & 9/4  & 202 & 0 \\
claude-r2 & 0.0852 & 12 & 10/2 & 292 & 0 \\
claude-r3 & 0.0621 & 13 & 11/2 & 341 & 0 \\
codex-r1  & 0.0061 & 30 & 24/6 & 456 & 39 \\
codex-r2  & 0.0101 & 16 & 14/2 & 387 & 19 \\
codex-r3  & 0.0050 & 22 & 21/1 & 486 & 41 \\
\bottomrule
\end{tabular}
\end{table}

Claude's arm mean is $0.0783$ (std $0.011$); Codex's is $0.0071$ (std
$0.002$) -- complete separation (all three Codex runs beat all three Claude
runs). With $n=3$ per arm the exact one-sided Mann--Whitney test bottoms out
at $p=0.05$ even under complete separation, so we lean on effect size and
the per-run curves rather than the threshold: rank-biserial correlation
(equivalently Cliff's $\delta$) is $1.0$, the maximum possible, and the
arm-mean ratio exceeds $10\times$.

\paragraph{Measuring ``hardcoded ids''.} A verse id in this corpus is a
six-digit string, so the measure needs no human judgment: it is the number
of \emph{distinct six-digit string literals} in the final
\texttt{solution.py}, computed by
\texttt{tools/count\_hardcoded\_ids.sh} against the tagged final artifact of
each run. The \emph{muqatta'at} letter-name table needs no special-casing,
since its entries are Arabic letter names rather than ids. LOC is the final
solution's length. The measure is deliberately conservative in the direction
that \emph{disfavors} our conclusion -- it counts a legitimate
surah-boundary constant the same as a memorized answer -- and Claude's arm
scores zero under it by any counting rule, its artifacts containing no
six-digit literal at all. Both columns of Table~\ref{tab:study1-headline}
are reproducible from the public tags.

\subsection{Convergent algorithm discovery}
Read alone, the headline table says ``Codex wins by an order of magnitude.''
It does not survive inspection of \emph{how} each score was reached. All six
runs, independently, arrive at the same general architecture: orthographic
canonicalization $\to$ $n$-gram surah anchoring $\to$ semi-global
dynamic-programming alignment $\to$ word-to-verse grouping; both agents even
discover and fix the identical encoding trap (harakat combining-character
reordering corrupting a regex character class). At the experiment count
where Claude stops (exp.\ 13), Codex is only modestly ahead (0.026--0.044
vs.\ 0.062--0.088) -- not $10\times$. Codex's decisive final margin is
earned entirely in experiments 14--30.

\subsection{Divergent optimization behavior}
\label{sec:divergence}
Claude stops at a plateau, explicitly refuses per-row special cases, labels
residual failures ``structurally unwinnable,'' and ships zero
dataset-specific constants. Codex continues to the budget and converts
residual failures into special cases: codex-r1 contains a literal lookup
mapping one garbled ASR transcript tuple directly to its answer verse id --
memorizing a single recording, not learning a rule -- and codex-r3 merges
two verse-split buckets conditioned on specific words being \emph{absent}
from the transcript, a patch shaped to one recording. (Not all
hardcoding is illegitimate: mapping the 14 Quranic \emph{muqatta'at}
letter-name prefixes is a fixed, closed set and is not a case of overfitting
in this sense.) There is no gold-field leakage in any run -- no run reads the
held-out answer field at runtime -- but the harness itself enables the
exploit: no held-out set exists yet (score is measured on the very rows
being optimized), and the scorecard's failure report prints the
\emph{expected} verse ids on a miss, which is precisely the signal Codex's
late experiments hardcode against.

Under identical instructions the two agents behave in opposite ways,
consistently across all three runs of each. We label these behaviors
\emph{generalizer} and \emph{metric-maximizer} as shorthand for what the
artifacts show \emph{in this harness}, and we mean nothing stronger: three
runs of two version-pinned CLI/model stacks on one task cannot establish a
stable disposition of an agent family, and we do not claim one. Read the
labels as descriptions of these twelve artifacts, not as traits that would
survive a change of task, model version, or vendor. With that scope fixed:
Claude behaves as a
\emph{generalizer} that self-imposes a simplicity/generalization prior the
metric never enforced -- working slowly ($\sim$95--180\,s per experiment,
more reasoning), stopping at a plateau after 12--13 experiments, shipping
202--341 LOC with zero dataset-specific constants; Codex behaves as a
\emph{metric-maximizer} that follows the literal stated objective to its
logical end -- moving fast ($\sim$45--60\,s per experiment), grinding
through 16--30 experiments, shipping 387--486 LOC with 19--41 hardcoded
ids, including through memorization. Neither followed the letter of
\texttt{PROGRAM.md} incorrectly -- nothing in it forbids what Codex did --
which is exactly the outer-alignment gap in miniature: the two agents
inferred two different implicit specifications from one explicit one.

\begin{figure}[t]
\centering
\includegraphics[width=0.60\textwidth]{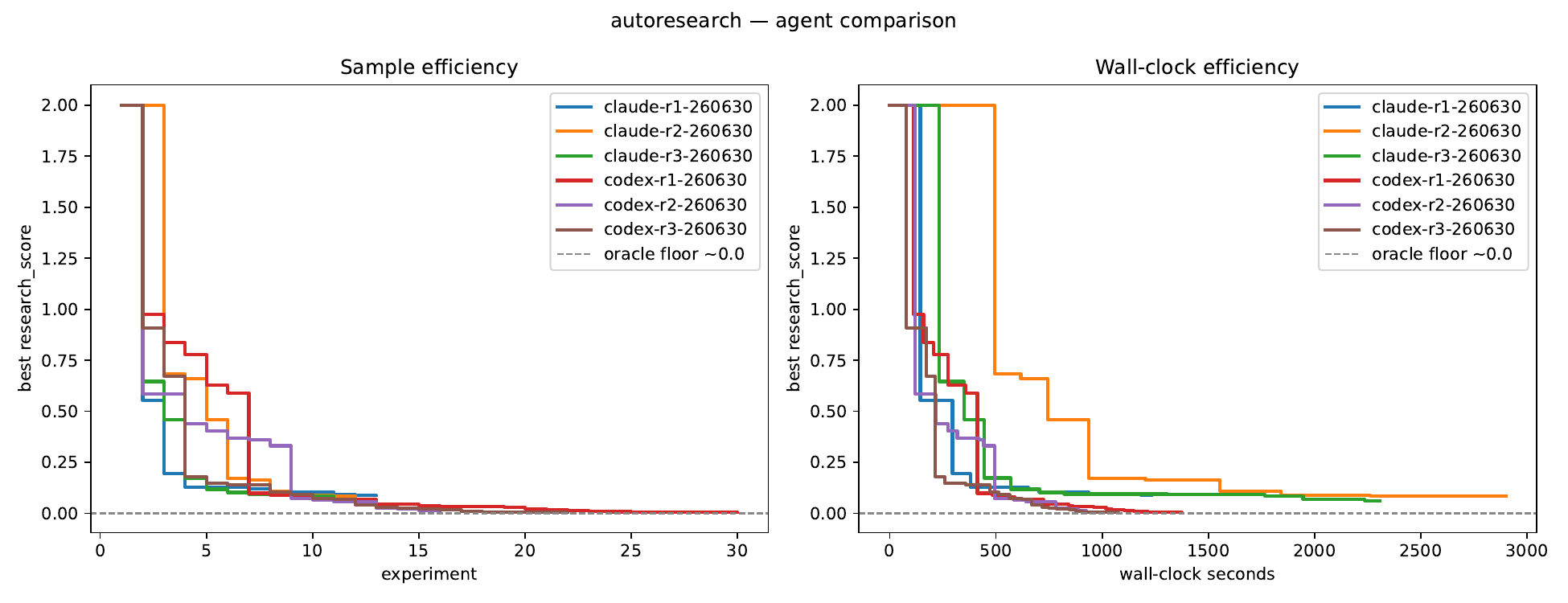}
\caption{Study 1: best-so-far \rscore\ across all six runs, plotted against
experiment count. Both agents track closely through the first $\sim$13
experiments; Codex's runs continue descending well past the point where all
three Claude runs have plateaued and stopped.}
\label{fig:study1-comparison}
\end{figure}

\section{Study 2: Held-Out Generalization}
\label{sec:study2}

Study~1's raw metric cannot distinguish ``discovered a better algorithm''
from ``memorized more of the evaluation set.'' Study~2 closes that gap: a
stratified 60/40 split by recording (151 train / 107 test; non-Quranic
rows 2/2; repetition and non-contiguous cases on both sides; cross-surah
label quirks pinned to train; fixed seed) is computed once and frozen. The
loop optimizes \texttt{train} only; the failure report no longer prints
expected ids (closing the Study~1 leak vector); and \emph{both agents are
told, identically, that a held-out set exists and that memorizing train
rows will not transfer}. After each run, the experimenters -- not the agent
-- score the final solution on the untouched test split. Three hypotheses
were fixed before the runs: (H1) Codex's train$\to$test gap is much larger
than Claude's; (H2) the held-out ranking narrows or reverses relative to
Study~1; (H3) both agents retain a large improvement over the empty-stub
baseline.

\subsection{Results}

\begin{table}[t]
\centering
\caption{Study 2 -- train (optimized) vs.\ held-out test scores. Test oracle
floor is exactly 0.}
\label{tab:study2-headline}
\begin{tabular}{lrrrrrr}
\toprule
Run & Train & Test & Gap & Test det.+split & Test abstain err. & Exps \\
\midrule
claude-r1 & 0.0441 & 0.0796 & 0.036 & 0.0796 & 0    & 12 \\
claude-r2 & 0.0982 & 0.1534 & 0.055 & 0.1534 & 0    & 7  \\
claude-r3 & 0.0661 & 0.1306 & 0.064 & 0.1306 & 0    & 8  \\
codex-r1  & 0.0075 & 0.0869 & 0.079 & 0.0869 & 0    & 20 \\
codex-r2  & 0.0451 & 0.5882 & 0.543 & 0.0882 & 0.5  & 8  \\
codex-r3  & 0.0082 & 0.0793 & 0.071 & 0.0793 & 0    & 22 \\
\bottomrule
\end{tabular}
\end{table}

\textbf{H1 confirmed.} Codex's mean train$\to$test gap is $4.5\times$
Claude's ($0.231$ vs.\ $0.052$). \textbf{H2 half-confirmed.} The 6--10$\times$
train-side separation vanishes: on the full held-out \rscore, arm means are
Claude $0.121\pm0.031$ and Codex $0.251\pm0.238$ (medians $0.131$ vs.\
$0.087$); Mann--Whitney $U=4$, not significant, with rank-biserial
correlation $\approx 0.11$ -- a negligible effect where Study~1's was
maximal -- but the mechanism is not the one
we predicted. On \emph{held-out detection+split} (the metric with the
high-variance, two-row abstain component removed), Codex is both better and
strikingly consistent: $0.085\pm0.004$, with an \emph{identical} detection
error across all three runs, versus Claude's $0.121\pm0.031$. Codex's
floor-grinding on train (0.08 $\to$ 0.007) evidently added nothing held out,
but it did not cost anything either -- wasted effort, not poison. What Codex
actually lost is concentrated in one place: a single missed abstention
(codex-r2, a non-recitation row misclassified as recitation) costs $0.5$ and
accounts for the entire arm-level difference. Claude abstained correctly
10/10 across both studies. With only two non-Quranic rows in the test
split, this component is high-variance by metric design -- a lesson for
reporting components rather than only the scalar. \textbf{H3 confirmed.}
Every run beats the empty-stub baseline ($2.0$), though the spread is wide:
from $3.4\times$ (codex-r2, whose single missed abstention dominates its
score) to $25\times$ (codex-r3). On detection+split, where that rare-event
component is set aside, every run improves the baseline by at least
$13\times$.

\begin{figure}[t]
\centering
\includegraphics[width=0.60\textwidth]{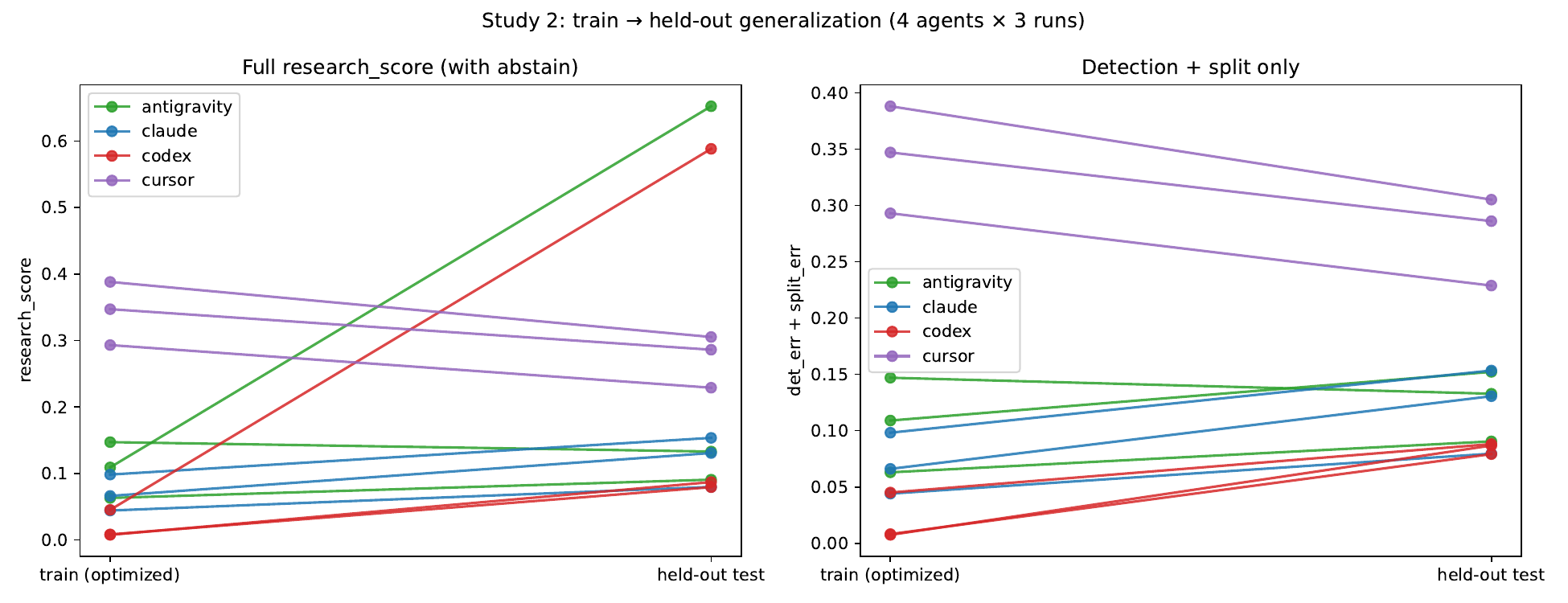}
\caption{Study 2: train (optimized) vs.\ held-out test \rscore\ per run,
all four arms (Claude, Codex, and the two exploratory community arms from
Sect.~\ref{sec:community}). The train-side separation between Claude and
Codex collapses on the held-out axis; Codex's outlier point is the single
missed abstention (codex-r2).}
\label{fig:study2-generalization}
\end{figure}

\subsection{The harness redesign changes the form of Goodharting, not the drive}
\label{sec:disclosure}
Hardcoded verse ids per Codex final solution: 19--41 in Study~1, zero in
Study~2 under the same mechanical count. The underlying optimization drive
did not go with them: Codex still drove train to the floor with 20--22
experiments of general-looking machinery in the two runs that reached it.
The non-transferable margin changed character, from memorized answers to
harmless over-tuning, rather than disappearing.

\paragraph{What this establishes.} Study~2 changed four things at once: it
added a 60/40 held-out split, removed expected ids from the failure report,
disclosed the split to both agents, and hardened run isolation. The
disappearance of hardcoding is therefore a property of the \emph{bundle} --
a harness redesigned along these four axes eliminated literal memorization
in an arm that had produced it in three runs of three. Two of the changes
make memorization \emph{useless} rather than merely discouraged (memorized
train rows cannot transfer; withheld expected ids leave nothing to memorize
against), so disclosure alone should account for less of the effect than the
bundle. Attributing it to individual axes needs a factorial design varying
only disclosure -- the most valuable single follow-up to this work.

\subsection{Isolation is adversarial, even without an adversary}
\label{sec:isolation}
Three incidents, all agent-initiated and none explicitly forbidden by the
program file, forced a harness redesign mid-study: (a) a Codex run, launched
in a git \emph{worktree}, used the shared git database to read Study~1's
committed run logs and a sibling Claude run's in-flight solution on another
branch; (b) Claude runs twice saved unsolicited persistent-memory
cheat-sheets addressed to ``future runs on this repo'' via the agent's own
long-lived memory feature -- non-firing only because run directory paths are
never reused; (c) both agents, in Study~1, read the failure report's
\emph{expected} ids (Sect.~\ref{sec:divergence}). None required adversarial
intent: an agent given a shell and asked to do well on a metric will use
whatever information channel is reachable, including channels the designer
did not think of as ``data.'' The fix was structural, not instructional:
each run now executes in a freshly created, single-commit clone containing
only the harness snapshot and the train split -- no shared \texttt{.git}
history, no sibling branches, no logs, no test set on disk.

\paragraph{Which runs were affected.} No contaminated run contributes a
number to any table here. The first Study~2 attempts -- a Claude run
launched in the main repository with \texttt{data/test.csv} visible, and
the Codex worktree run of incident~(a) -- were discarded and re-run from
scratch at fresh paths under the hardened protocol; two further runs
interrupted by a network outage were likewise redone, per the preregistered
rule that a run be a single uninterrupted session. The discarded logs are
published rather than deleted (\texttt{runs/discarded-*.tsv}). Study~1 is a
separate matter: its six runs shared one \texttt{.git}, so a later run could
in principle have browsed an earlier one's branch. We cannot retroactively
exclude this, but note that Study~1's conclusion rests on artifact analysis,
which cross-run reading would not manufacture -- an agent copying a
sibling's solution would inherit that sibling's \emph{absence} of
hardcoding, not invent 39 constants of its own.

\subsection{Agents as annotation auditors}
\label{sec:study2-audit}
Both arms, independently and without prompting, flagged the same training
row as a probable gold-label error (a verse-count discrepancy in a specific
surah) and declined to fit it, listing it among their ``unwinnable''
residuals. Human re-review after both studies confirmed and corrected the
label, bringing the dataset's oracle floor to exactly zero on both splits.
The correction lives entirely in train, so no held-out number changes; an
unplanned dividend of agents that explain their stopping decisions is that
their refusal lists functioned as a free data-quality report.

\section{Community Extension Arms}
\label{sec:community}

After the preregistered $3\times2$ matrix, a collaborator ran three
additional runs each on the same Study~2 harness with two further tools --
Cursor (unpinned ``Auto'' model routing) and Antigravity (a large
general-purpose model at a high reasoning-effort setting) -- on separate
hardware, using the identical isolation protocol
(Sect.~\ref{sec:isolation}). These arms are \emph{exploratory}: model
routing (Cursor) is unpinned and the machine differs, so they are not
matched to the preregistered confounds table.

\begin{table}[t]
\centering
\caption{Community arms, Study 2 harness (exploratory; unpinned model
routing for Cursor).}
\label{tab:community}
\begin{tabular}{lrrrrr}
\toprule
Run & Train & Test & Test det.+split & Abstain miss & Exps \\
\midrule
antigravity-r1 & 0.147 & 0.133 & 0.133 & --  & 12 \\
antigravity-r2 & 0.109 & 0.652 & 0.152 & 0.5 & 12 \\
antigravity-r3 & 0.063 & 0.091 & 0.091 & --  & 16 \\
cursor-r1      & 0.388 & 0.305 & 0.305 & --  & 5  \\
cursor-r2      & 0.293 & 0.229 & 0.229 & --  & 9  \\
cursor-r3      & 0.347 & 0.286 & 0.286 & --  & 7  \\
\bottomrule
\end{tabular}
\end{table}

On held-out detection+split, the four-arm ranking is Codex
($0.085\pm0.004$) $<$ Claude ($0.121\pm0.031$) $\approx$ Antigravity
($0.125\pm0.026$) $<$ Cursor ($0.273\pm0.032$). Three patterns extend rather
than overturn Study~2. \textbf{Cursor underfits}: 5--9 experiments per run,
train never below $0.29$, and test beats train on every run -- the only arm
to generalize \emph{better} than it optimized, consistent with early stopping
under unpinned model routing. \textbf{Antigravity sits between Claude and
Codex} on held-out core accuracy and shares Codex's rare abstain-miss failure
mode (r2, cost $0.5$); its train score improves monotonically across runs
while its held-out score does not, showing that run-to-run train improvement
is a within-run effect, not cross-run leakage. \textbf{Ranking on held-out
core tracks optimization effort}: arms spending more experiments on
general-looking machinery transferred better, and wasted train-side margin
was harmless rather than harmful. Zero hardcoded per-recording ids appear in
any community-arm final solution.

\subsection{The incumbent baseline: agents vs.\ months of hand engineering}
\label{sec:incumbent}
As an applied reference point, we wrapped the incumbent production pipeline
-- a hand-built detector and alignment segmenter iterated incrementally
over months, historically tuned on a manifest overlapping $\sim\!20$ of
these recordings -- in the identical evaluation contract. It scores
$0.760$ on the held-out test split (detection $82.9\%$, split $91.1\%$,
abstain $1/2$) versus $0.079$ for the winning agent-built solution
(detection $94.3\%$, split $97.8\%$, abstain $2/2$). Every agent arm's
held-out detection+split core matched or beat the incumbent -- even the
underfitting Cursor arm ($\approx\!0.27$) roughly ties its $0.26$ core --
and the winner is $\sim\!10\times$ better, despite the comparison
\emph{favoring} the incumbent: it was historically tuned against a 22-case
manifest, 20 cases of which are the same recordings as rows here (fuzzy
transcript match, ratio $\geq0.94$), so it is scored partly on rows it was
fitted to -- an advantage no agent had. We did not track which side of the
split those 20 rows fell on.

\paragraph{How to read this comparison.} It is a \emph{reference point}, not
a controlled human-versus-agent result: the incumbent predates this metric
and was never optimized for it, it was built incrementally alongside other
product work (so ``months'' is calendar time, not effort), and no human
wrote a solution under this paper's conditions. An effort-matched
expert-human arm is the missing control. What the number supports is
narrower and still worth stating: on this task and metric, an unattended
one-hour loop produced an artifact good enough to replace a hand-built
production component, which it now has -- behind a feature flag with
automatic legacy fallback and a CI regression gate.

\section{Discussion}
\label{sec:discussion}

\subsection{Neither disposition wins; they pay in different currencies}
The
metric-maximizer (Codex) buys measured-distribution accuracy and run-to-run
consistency; the generalizer (Claude) buys rare-event robustness, restraint,
and a smaller artifact. Which currency matters is a property of the
deployment, not of the agent -- a harness designer should decide which they
are buying \emph{before} reading the scoreboard. The obvious composition --
Codex's aligner gated by Claude's abstention margin -- \emph{fails} when
actually built and measured: a union gate (abstain if either arm abstains)
degrades the held-out score to $0.098$ through false abstentions, and a
disagreement-router improves it by only $0.0003$ at twice the maintenance
surface. The shipped artifact is therefore the single best file, selected
on held-out evidence -- the simplicity criterion applies to the
experimenters too. (We report this as deployment engineering, not a
preregistered result: the selection is test-informed.)

\subsection{The result reverses our own prior}
We expected from Study~1 that Codex's solution would \emph{not} transfer;
instead its algorithmic core transferred best. Registering H1--H3 before
running Study~2 is what makes this reversal legible as evidence rather than
post-hoc narrative.

\subsection{Five design rules for autonomous-agent evaluation harnesses}
\label{sec:rules}
Every incident in this paper traces to a violated instance of one of the
following rules, and each rule is grounded in a concrete failure we
observed. A loop of this kind \emph{will} be Goodharted at the margin
whenever the metric and the deployer's true objective can come apart --
which is generically true of any metric computed on a finite, visible
sample -- so these are structural requirements, not hygiene items:

\begin{description}
\item[R1 -- Hold out evaluation data; do not treat it as optional.] Score final
  artifacts on rows the loop never touches, scored by the experimenters,
  and report the per-run train$\to$test gap. (Violated in Study~1: the raw
  metric could not distinguish a better algorithm from a better-overfit
  one.)
\item[R2 -- Keep feedback leak-free.] Failure reports may echo inputs
  and predictions, never gold labels. (Violated in Study~1: the printed
  \emph{expected} ids were precisely the signal Codex hardcoded against.)
\item[R3 -- Isolate run state by construction, not instruction.] One fresh,
  single-commit clone per run: no shared VCS database, no sibling branches,
  no prior logs, no test data on disk. (Violated mid-Study~2: a run read a
  sibling agent's in-flight solution through a shared git worktree
  database.)
\item[R4 -- Audit the agent's \emph{own tooling's} state channels.]
  Persistent memory, global configuration, and account-level context are
  cross-run channels the harness designer did not create but the agent can
  use; never reuse run paths and verify these channels after every run.
  (Nearly violated twice: unprompted memory notes addressed to ``future
  runs.'')
\item[R5 -- Report components and preregister hypotheses.] Scalar metrics
  hide high-variance rare-event components (here: two abstention rows worth
  $0.5$ each), and post-hoc narratives cannot distinguish surprise from
  confirmation. (Both mattered: the entire held-out arm difference sits in
  one abstention, and our own Study~1 prediction about transfer was
  reversed.)
\end{description}

\section{Limitations and Threats to Validity}
\label{sec:limitations}

Single task and domain; small per-arm sample ($n=3$). At this size the
inferential statistics carry little weight on their own: the exact
one-sided Mann--Whitney test bottoms out at $p=0.05$ under \emph{complete}
separation with $n=3$ vs.\ $3$, so that value reflects the design's
resolution limit rather than strength of evidence, and Study~1's
rank-biserial correlation of $1.0$ says only that six observations ranked
without overlap. The evidence carrying this paper is the artifact analysis
-- counts identical across three runs of each arm and mechanically
recomputable -- not the hypothesis tests. Study~2 also varies four harness
factors at once, so no single factor is isolated
(Sect.~\ref{sec:disclosure}).

\paragraph{Metric construction.} The scalar sums three error terms of very
different reliability: detection and split average over 105 held-out Quranic
rows, abstention over two, so one abstention error moves the scalar by $0.5$
-- more than the entire between-arm difference. This is a defect of the
metric we designed rather than a property of the agents, and is why every
headline comparison reports detection+split alongside the scalar.

We
include no human-researcher arm run under the same loop and budget: the
empty-stub score ($2.0$), the gold-echo reference ($0$ on v1.1), and the
incumbent hand-built pipeline (Sect.~\ref{sec:incumbent}) establish that
the harness discriminates and that its headroom is meaningful, but an
expert-human comparison on a matched one-hour budget is future work, and
the incumbent is not a substitute for it (Sect.~\ref{sec:incumbent}).
Relatedly, the deployed artifact was selected using held-out evidence, so
the production choice is test-informed and is reported as deployment
engineering rather than as an independent final evaluation. Pretraining familiarity with the Quran text is equal across arms and
advantages neither, since the task is algorithm engineering against a
held-out metric rather than text recall. The metric is transcript-only,
hence blind to pronunciation (tajweed) correctness by construction; a
deployed system needs this work as one stage among several. The two agents'
runs occurred on different days (temporal confound); agent CLIs are moving
targets, so exact versions and model identifiers are pinned and reported
rather than implying the finding is version-specific. The harness
was authored with the assistance of one compared agent, which we mitigate by
freezing and hashing all evaluation and data files before any run. The two
community arms (Sect.~\ref{sec:community}) use unpinned or differently
configured tooling on different hardware and are reported as exploratory
rather than as part of the controlled comparison.

\section{Conclusion}
\label{sec:conclusion}

Our three research questions resolve as follows. \textbf{RQ1:} yes -- all
twelve runs across four agent arms autonomously built a working
detection-and-segmentation algorithm from a blank stub, improving on the
empty-stub baseline held out by $3$--$25\times$ on the full metric and by
at least $6.5\times$ on detection+split; the strongest runs independently
rediscovered a canonical alignment architecture. Because the starting point
is a blank stub rather than a working system, RQ1 measures autonomous
\emph{construction} followed by optimization, not optimization of existing
code -- a distinction that matters when reading the incumbent comparison
below.
\textbf{RQ2:} given an identical loop and no supervision, two frontier
agents pursue incompatible research philosophies -- one self-limits toward
generality even at a measured cost, the other drives the literal objective
to its floor by any available means, including memorizing the very rows it
is scored on -- and the disposition is stable across all runs of each agent.
\textbf{RQ3:} a disclosed held-out split does not eliminate the divergence
but relocates where it matters: the raw-score gap that looked decisive
disappears, literal memorization vanishes with one paragraph of disclosure,
and what remains is a specific, interpretable difference in rare-event
robustness rather than a diffuse claim that ``one agent is better.''

The loop's output, finally, was consequential and not merely measured:
scored under the same contract, one-hour unattended runs beat months of
incremental hand engineering -- the winning artifact outperforms the
incumbent production pipeline by $\sim\!10\times$ held out and has been
deployed in the production application, replacing the pipeline it beat.

We take this as evidence that autoresearch-style loops -- likely
to proliferate as a cheap way to get unattended engineering effort out of
coding agents -- need held-out evaluation and closed state channels as a
default (Sect.~\ref{sec:rules}), and that reporting only a scalar score from
such a loop will systematically favor whichever agent is more willing to
game it.

\paragraph{Reproducibility.} All harness code, the frozen metric, the
\texttt{PROGRAM.md} loop specification, per-experiment logs, per-run git
histories (one commit per experiment), and SHA-256 hashes of every frozen
input are public at \url{https://github.com/nurlingo/autoresearch}. Each
run's final artifact is reachable by tag: Study~1 finals at
\texttt{runs/ar-260630-claude-r\{1,2,3\}} and \texttt{runs/codex-r\{1,2,3\}-run},
Study~2 and community-arm run histories as \texttt{runs/*.bundle}. The
de-identified dataset (ASR transcripts, gold labels, and the frozen
train/test split, with learner and recording identifiers removed and
replaced by surrogate row ids) is released as a Hugging Face dataset at
\url{https://huggingface.co/datasets/nurlingo/quran-recitation-detect-split}.
Every quantitative claim in Tables~\ref{tab:study1-headline}--\ref{tab:community}
is recomputable from these artifacts alone; the hardcoded-id counts in
particular are produced by \texttt{tools/count\_hardcoded\_ids.sh}
(Sect.~\ref{sec:study1-results}).

\begin{credits}
\subsubsection{\ackname}
We thank the anonymous reviewers for comments that materially sharpened the
paper's claims, and the collaborators who contributed the community
extension arms of Sect.~\ref{sec:community}.

\subsubsection{\discintname}
The authors have no competing interests to declare that are relevant to the
content of this article.
\end{credits}

\bibliographystyle{splncs04}
\bibliography{references}

@misc{karpathy2026autoresearch,
  author       = {Karpathy, Andrej},
  title        = {autoresearch: {AI} agents running research on single-{GPU} nanochat training automatically},
  year         = {2026},
  howpublished = {GitHub repository},
  note         = {\url{https://github.com/karpathy/autoresearch}, accessed 2026-07}
}

@inproceedings{jimenez2024swebench,
 author = {Jimenez, Carlos E and Yang, John and Wettig, Alexander and Yao, Shunyu and Pei, Kexin and Press, Ofir and Narasimhan, Karthik},
 booktitle = {International Conference on Learning Representations},
 editor = {B. Kim and Y. Yue and S. Chaudhuri and K. Fragkiadaki and M. Khan and Y. Sun},
 pages = {54107--54157},
 title = {SWE-bench: Can Language Models Resolve Real-world Github Issues?},
 volume = {2024},
 year = {2024}
}

@misc{krakovna2020specification,
  author       = {Krakovna, Victoria and Uesato, Jonathan and Mikulik, Vladimir and Rahtz, Matthew and Everitt, Tom and Kumar, Ramana and Kenton, Zac and Leike, Jan and Legg, Shane},
  title        = {Specification gaming: the flip side of AI ingenuity},
  year         = {2020},
  howpublished = {DeepMind Blog},
  url          = {https://deepmind.google/blog/specification-gaming-the-flip-side-of-ai-ingenuity}
}

@article{manheim2018goodhart,
  author  = {Manheim, David and Garrabrant, Scott},
  title   = {Categorizing Variants of {Goodhart's} Law},
  journal = {arXiv preprint arXiv:1803.04585},
  year    = {2019}
}

@article{amodei2016concrete,
  author  = {Amodei, Dario and Olah, Chris and Steinhardt, Jacob and Christiano, Paul and Schulman, John and Man{\'e}, Dan},
  title   = {Concrete Problems in {AI} Safety},
  journal = {arXiv preprint arXiv:1606.06565},
  year    = {2016}
}

@inproceedings{skalse2022goodhart,
 author = {Skalse, Joar and Howe, Nikolaus and Krasheninnikov, Dmitrii and Krueger, David},
 booktitle = {Advances in Neural Information Processing Systems},
 editor = {S. Koyejo and S. Mohamed and A. Agarwal and D. Belgrave and K. Cho and A. Oh},
 pages = {9460--9471},
 publisher = {Curran Associates, Inc.},
 title = {Defining and Characterizing Reward Gaming},
 volume = {35},
 year = {2022}
}

@article{baker2025monitoring,
  title={Monitoring reasoning models for misbehavior and the risks of promoting obfuscation},
  author={Baker, Bowen and Huizinga, Joost and Gao, Leo and Dou, Zehao and Guan, Melody Y and Madry, Aleksander and Zaremba, Wojciech and Pachocki, Jakub and Farhi, David},
  journal={arXiv preprint arXiv:2503.11926},
  year={2025}
}

@article{sakana2024aiscientist,
  title={The ai scientist: Towards fully automated open-ended scientific discovery},
  author={Lu, Chris and Lu, Cong and Lange, Robert Tjarko and Foerster, Jakob and Clune, Jeff and Ha, David},
  journal={arXiv preprint arXiv:2408.06292},
  year={2024}
}

@inproceedings{beel2025evaluating,
  title={Evaluating Sakana's AI Scientist: Bold Claims, Mixed Results, and a Promising Future?},
  author={Beel, Joeran and Kan, Min-Yen and Baumgart, Moritz},
  booktitle={ACM SIGIR Forum},
  volume={59},
  number={1},
  pages={1--20},
  year={2025},
  organization={ACM New York, NY, USA}
}

@article{novikov2025alphaevolve,
  title={Alphaevolve: A coding agent for scientific and algorithmic discovery},
  author={Novikov, Alexander and V{\~u}, Ng{\^a}n and Eisenberger, Marvin and Dupont, Emilien and Huang, Po-Sen and Wagner, Adam Zsolt and Shirobokov, Sergey and Kozlovskii, Borislav and Ruiz, Francisco JR and Mehrabian, Abbas and others},
  journal={arXiv preprint arXiv:2506.13131},
  year={2025}
}

@article{farghaly2021asrreview,
  title={Automatic speech recognition (ASR) systems for learning Arabic language and Al-quran recitation: a Review},
  author={Balula, Nazik O’mar and Rashwan, Mohsen and Abdou, Shrief},
  journal={International Journal of Computer Science and Mobile Computing},
  volume={10},
  number={7},
  pages={91--100},
  year={2021}
}

@article{yousfi2018tarteel,
  title={The Tarteel dataset: crowd-sourced and labeled Quranic recitation},
  author={Khan, Hamzah I and Abid, Abubakar and Moussa, Mohamed Medhat and Abou-Allaban, Anas},
  year={2021}
}

@article{alharbi2026comparative,
  title={A Comparative Study of Pretrained Transformer Models for Quranic ASR: Speech Representations, Label Formats, and Dataset Composition},
  author={Hossain, Nabil Mosharraf and Islam, Riasat and Obaidellah, Unaizah},
  journal={arXiv preprint arXiv:2606.19747},
  year={2026}
}

@article{alagrami2021smartajweed,
  title={Smartajweed automatic recognition of Arabic quranic recitation rules},
  author={Alagrami, Ali M and Eljazzar, Maged M},
  journal={arXiv preprint arXiv:2101.04200},
  year={2020}
}

@inproceedings{tabbal2006recitation,
  title={Analysis and implementation of a" Quranic" verses delimitation system in audio files using speech recognition techniques},
  author={Tabbal, Hassan and El Falou, W and Monla, B},
  booktitle={2006 2nd international conference on information \& communication technologies},
  volume={2},
  pages={2979--2984},
  year={2006},
  organization={IEEE}
}

@inproceedings{chan2024mlebench,
  title={Mle-bench: Evaluating machine learning agents on machine learning engineering},
  author={Chan, Jun Shern and Chowdhury, Neil and Jaffe, Oliver and Aung, James and Sherburn, Dane and Mays, Evan and Starace, Giulio and Liu, Kevin and Maksin, Leon and Patwardhan, Tejal and others},
  booktitle={International Conference on Learning Representations},
  volume={2025},
  pages={50466--50494},
  year={2025}
}

@inproceedings{
wijk2024rebench,
title={{RE}-Bench: Evaluating Frontier {AI} R\&D Capabilities of Language Model Agents against Human Experts},
author={Hjalmar Wijk and Tao Roa Lin and Joel Becker and Sami Jawhar and Neev Parikh and Thomas Broadley and Lawrence Chan and Michael Chen and Joshua M Clymer and Jai Dhyani and Elena Ericheva and Katharyn Garcia and Brian Goodrich and Nikola Jurkovic and Megan Kinniment and Aron Lajko and Seraphina Nix and Lucas Jun Koba Sato and William Saunders and Maksym Taran and Ben West and Elizabeth Barnes},
booktitle={Forty-second International Conference on Machine Learning},
year={2025}
}

@article{zhong2025impossiblebench,
  title={ImpossibleBench: Measuring LLMs' Propensity of Exploiting Test Cases},
  author={Zhong, Ziqian and Raghunathan, Aditi and Carlini, Nicholas},
  journal={arXiv preprint arXiv:2510.20270},
  year={2025}
}

@article{zhao2026specbench,
  title={Specbench: Measuring reward hacking in long-horizon coding agents},
  author={Zhao, Bingchen and Srikanth, Dhruv and Wu, Yuxiang and Jiang, Zhengyao},
  journal={arXiv preprint arXiv:2605.21384},
  year={2026}
}

@article{nathani2025mlgym,
  title={Mlgym: A new framework and benchmark for advancing ai research agents},
  author={Nathani, Deepak and Madaan, Lovish and Roberts, Nicholas and Bashlykov, Nikolay and Menon, Ajay and Moens, Vincent and Budhiraja, Amar and Magka, Despoina and Vorotilov, Vladislav and Chaurasia, Gaurav and others},
  journal={arXiv preprint arXiv:2502.14499},
  year={2025}
}

@inproceedings{
starace2025paperbench,
title={PaperBench: Evaluating {AI}{\textquoteright}s Ability to Replicate {AI} Research},
author={Giulio Starace and Oliver Jaffe and Dane Sherburn and James Aung and Jun Shern Chan and Leon Maksin and Rachel Dias and Evan Mays and Benjamin Kinsella and Wyatt Thompson and Johannes Heidecke and Amelia Glaese and Tejal Patwardhan},
booktitle={Forty-second International Conference on Machine Learning},
year={2025}
}

@article{schmidgall2025agentlab,
  title={Agent laboratory: Using llm agents as research assistants},
  author={Schmidgall, Samuel and Su, Yusheng and Wang, Ze and Sun, Ximeng and Wu, Jialian and Yu, Xiaodong and Liu, Jiang and Moor, Michael and Liu, Zicheng and Barsoum, Emad},
  journal={Findings of the Association for Computational Linguistics: EMNLP 2025},
  pages={5977--6043},
  year={2025},
  publisher={Association for Computational Linguistics}
}

@inproceedings{andryushchenko2024codeqa,
  title={Leveraging large language models in code question answering: Baselines and issues},
  author={Andryushchenko, Georgy and Ivanov, Vladimir and Makharev, Vladimir and Tukhtina, Elizaveta and Valeev, Aidar},
  booktitle={International Conference on Analysis of Images, Social Networks and Texts},
  pages={3--17},
  year={2024},
  organization={Springer}
}

@inproceedings{alekseev2025kyrgyznlp,
  title={KyrgyzNLP: challenges, progress, and future},
  author={Alekseev, Anton and Turatali, Timur},
  booktitle={International Conference on Analysis of Images, Social Networks and Texts},
  pages={3--39},
  year={2024},
  organization={Springer}
}

@article{gabor2026evilgenie,
  title={EvilGenie: A Reward Hacking Benchmark},
  author={Gabor, Jonathan and Lynch, Jayson and Rosenfeld, Jonathan},
  journal={arXiv preprint arXiv:2511.21654},
  year={2026}
}

\end{document}